\input{aipcheck.tex}
\newcommand\sps{\space\space\space\space}
\documentclass{aipproc}

\layoutstyle{6x9}

\begin{document}

\title {Deeply Virtual Compton Scattering with Positron Beams at Jefferson Lab}


\author{Volker D. Burkert}{
  address={Jefferson Lab, 26000 Jefferson Avenue, Newport News, Virginia, USA},
  email={burkert@jlab.org}}

\begin{abstract}
 A brief discussion of the DVCS program at the Jefferson Lab 12 GeV energy upgrade 
 is given. Emphasis is on what can be learned from using both polarized electron and 
 polarized positron beams in conjunction with polarized nucleon targets.
\end{abstract}

\maketitle

\section{Introduction}
\label{intro}
The challenge of understanding nucleon electromagnetic structure still 
continues after more than five decades of experimental scrutiny. From the initial 
measurements of elastic form factors to the accurate determination of 
parton distributions through deep inelastic scattering, the
experiments have increased in statistical and systematic accuracy.  Only 
during the past decade it was realized that the parton distribution functions
represent special cases of a more general, much more powerful, way to 
characterize the structure of the nucleon, the generalized parton 
distributions (GPDs). For recent reviews see~\cite{mdiehl,radyush}.

\begin{figure}[h]
\includegraphics[height=0.3\textheight]{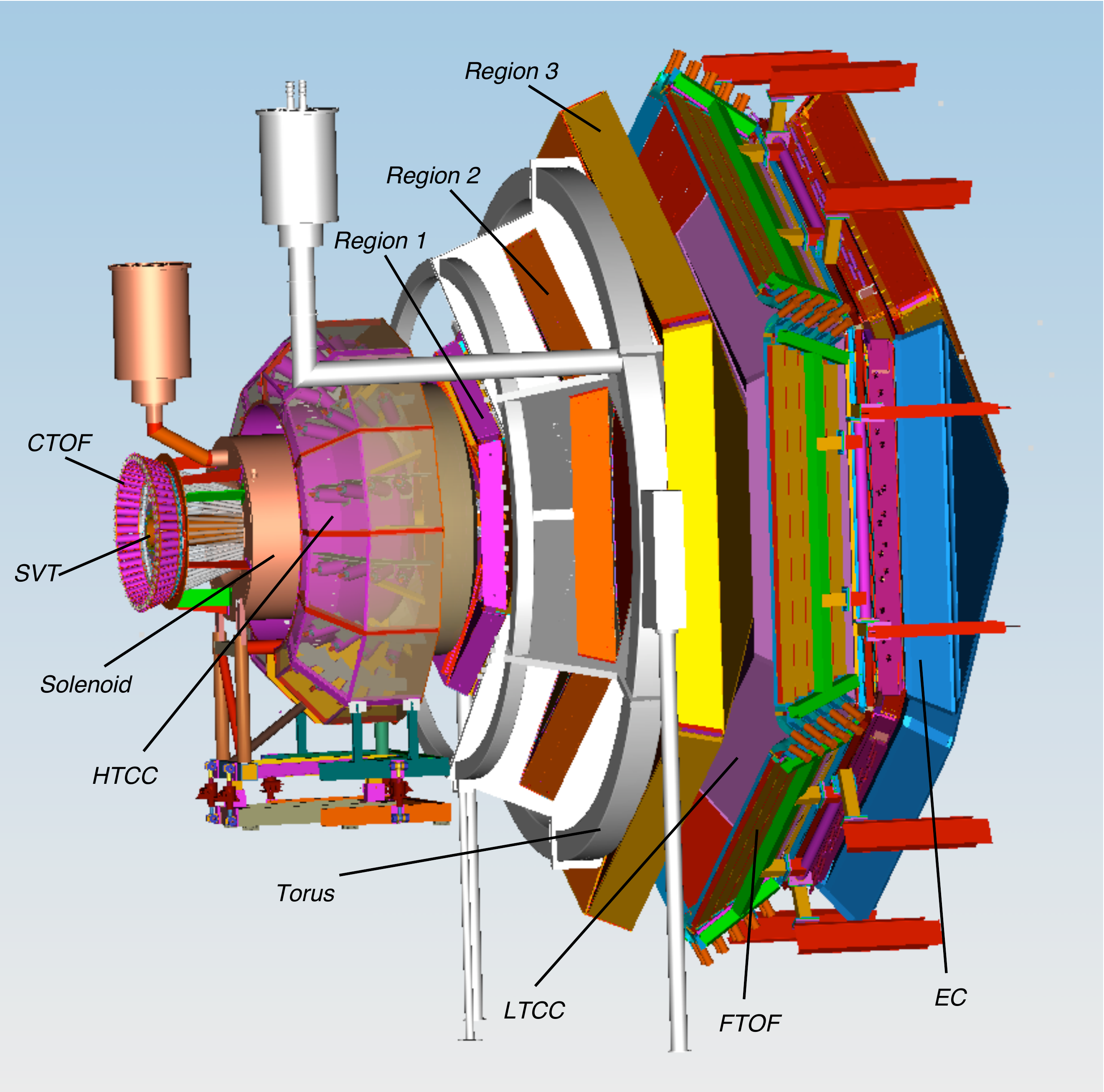}
\caption{The {\tt\sl CLAS12} detector is currently under construction to explore deeply virtual exclusive 
processes such as DVCS at the Jefferson Lab 12 GeV upgrade.}
\label{fig:clas12}
\end{figure}

 The GPDs are the Wigner quantum phase space 
distribution of quarks in the nucleon describing the 
simultaneous distribution of particles with respect to both position and 
momentum in a quantum-mechanical system.
In addition to the information about the spatial density 
and momentum density, these functions reveal the 
correlation of the spatial and momentum distributions, {\it i.e.} how the 
spatial shape of the nucleon changes when probing quarks of 
different momentum fraction of he nucleon.

The concept of GPDs has led to completely new methods of ``spatial imaging''
of the nucleon in the form of (2+1)-dimensional tomographic images, with 2 spatial
dimensions and 1 dimension in momentum~\cite{burkardt2003,ji2003}. The second moments 
of GPDs are related to 
form factors that allow us to quantify how the orbital motion of quarks in the nucleon contributes to the 
nucleon spin, and how the quark masses and the forces on quarks are distributed in 
transverse space, a question of crucial importance for our understanding of 
the dynamics underlying nucleon structure. 

The four leading twist GPDs $H$, $\tilde{H}$, $E$, and $\tilde{E}$, 
depend on the 3 variable $x$, $\xi$, and $t$, where $x$ is the longitudinal
momentum fraction of the struck quark, $\xi$ is the longitudinal momentum transfer
to the quark ($\xi \approx x_B/(2-x_B)$), and $t$ is the invariant 
4-momentum transfer to the proton.
The mapping of the nucleon GPDs, and a detailed understanding of the
spatial quark and gluon structure of the nucleon, have been widely 
recognized as key objectives of nuclear physics of the 
next decades. This requires a comprehensive program, combining results
of measurements of a variety of processes in electron--nucleon 
scattering with structural information obtained from theoretical studies, 
as well as with expected results from future lattice QCD simulations. The {\tt \sl CLAS12} 
detector, shown in Fig.~\ref{fig:clas12}, is currently under construction to pursue such 
an experimental program at the Jefferson Lab 12 GeV upgrade.

\begin{figure}[b]
\includegraphics[height=0.2\textheight]{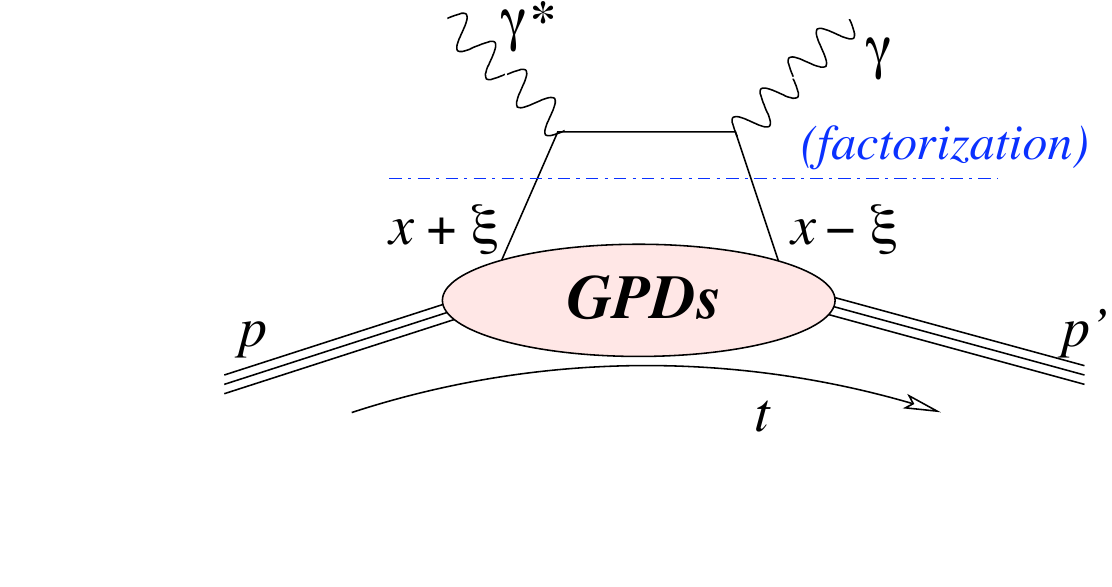}
\caption{Leading order contributions to the production of high energy single photons from protons. 
The DVCS handbag diagram contains the information on the unknown GPDs.}
\label{fig:handbag}
\end{figure}

\section {Accessing GPDs in DVCS}
The most direct way of accessing GPDs at lower energies is through the measurement 
of Deeply Virtual Compton Scattering (DVCS) in a kinematical domain where the 
so-called handbag diagram shown in Fig. ~\ref{fig:handbag} makes the dominant contributions. 
However, in DVCS as in other deeply virtual reactions, the GPDs do not appear directly in the 
cross section, but in convolution integrals, e.g. 
\begin{equation}
 \int_{-1}^{+1}{{H^q(x,\xi,t)dx}\over {x - \xi + i\epsilon}} = \int_{-1}^{+1}{{H^q(x,\xi,t)dx}\over {x - \xi }} + i\pi H^q(\xi,\xi,t)~,
 \end{equation}
 where the first term on the r.h.s. corresponds to the real part and the second term to the imaginary part of the 
scattering amplitude. The superscript $q$ indicates that GPDs depend on the quark flavor. From the 
above expression it is obvious that GPDs, in general, can not be accessed directly in measurements. However, 
in some kinematical regions the Bethe-Heitler (BH) process where high energy photons are 
emitted from the incoming and scattered electrons, can be important. Since the BH amplitude is  
purely real, the interference with the DVCS amplitude isolates the imaginary part of the DVCS 
amplitude. The interference of the two 
processes offers the unique possibility to determine GPDs directly at the singular kinematics $x=\xi$. 
At other kinematical regions 
a deconvolution of the cross section is required to determine the kinematic dependencies of the 
GPDs. It is therefore important to obtain all possible independent information that will aid in extracting 
information on GPDs. The interference 
terms for polarized beam $I_{LU}$, longitudinally polarized target $I_{UL}$, transversely (in scattering plane) 
polarized target $I_{UT}$, and perpendicularly (to scattering plane) polarized target $I_{UP}$ are given 
by the expressions: 
\begin{equation}
I_{LU} \sim  \sqrt{\tau^\prime} [F_1 H + \xi (F_1+F_2) \tilde{H} + \tau F_2 E] 
\end{equation}
\begin{equation}
I_{UL} \sim  \sqrt{\tau^\prime} [F_1 \tilde{H} + \xi (F_1 + F_2) H + (\tau F_2 - \xi F_1)\xi \tilde{E}]
\end{equation}
\begin{equation}
I_{UP} \sim {\tau}[F_2 H - F_1 E + \xi (F_1 + F_2)\xi \tilde{E}
\end{equation}
\begin{equation}
I_{UT} \sim  {\tau}[F_2 \tilde{H} + \xi (F_1 + F_2) E - (F_1+ \xi F_2) \xi \tilde{E}]
\end{equation}
\noindent
where $\tau = -t/4M^2$, $\tau^\prime = (t_0 - t)/4M^2$. By measuring all 4 combinations of interference 
terms one can separate all 4 leading twist GPDs at the specific kinematics $x=\xi$. Experiments at 
JLab using 4 to 6 GeV electron beams have
been carried out with polarized beams ~\cite{step01,munoz06,girod08,gavalian09} and with longitudinal
target~\cite{chen06}, showing the feasibility of such measurements at relatively low beam energies, and
their sensitivity to the GPDs. In the following sections we discuss what information may be gained 
by employing both electron and positron beams in deeply virtual photon production.

\subsection{Differential cross section for polarized leptons}
The structure of the differential cross section for polarized beam and unpolarized target is given by:
\begin{equation}
\sigma_{\vec{e}p\rightarrow e\gamma p} = \sigma_{BH} + e_\ell \sigma_{INT} + P_\ell e_\ell \tilde\sigma_{INT} + \sigma_{VCS} + P_\ell \tilde\sigma_{VCS} 
\end{equation}
\noindent 
where $\sigma$ is even in azimuthal angle $\phi$, and $\tilde\sigma$ is odd in $\phi$. The interference terms
 $\sigma_{INT} \sim \rm{Re} {\it A}_{\gamma^*N\rightarrow \gamma N} $ and   
 $\tilde\sigma_{INT} \sim \rm{Im} {\it A}_{\gamma^*N\rightarrow \gamma N} $ are 
 the real and imaginary parts, respectively of the Compton amplitude. Using polarized 
 electrons the combination $-\tilde\sigma_{INT} + \tilde\sigma_{VCS}$ can be determined 
 by taking the difference of the beam helicities. The electron-positron charge difference for 
 unpolarized beams determines $\sigma_{INT}$. For fixed beam polarization and taking the 
 electron-positron difference one can extract the combination $P_\ell\tilde\sigma_{INT} + \sigma_{INT}$. 
If only a polarized electron beam is available one can separate $\tilde\sigma_{INT}$ 
from $\tilde\sigma_{VCS}$ using the Rosenbluth technique. This requires measurements at 
two significantly different beam energies which reduces the kinematical coverage that can be achieved 
with this method. With polarized electrons and polarized positrons 
both $\sigma_{INT}$ can be determined and  $\tilde\sigma_{INT}$ can be 
separated from $\tilde\sigma_{VCS}$ in the full kinematic range available at the 
maximum beam energy.   
\begin{figure}[htb]
  \includegraphics[height=.24\textheight]{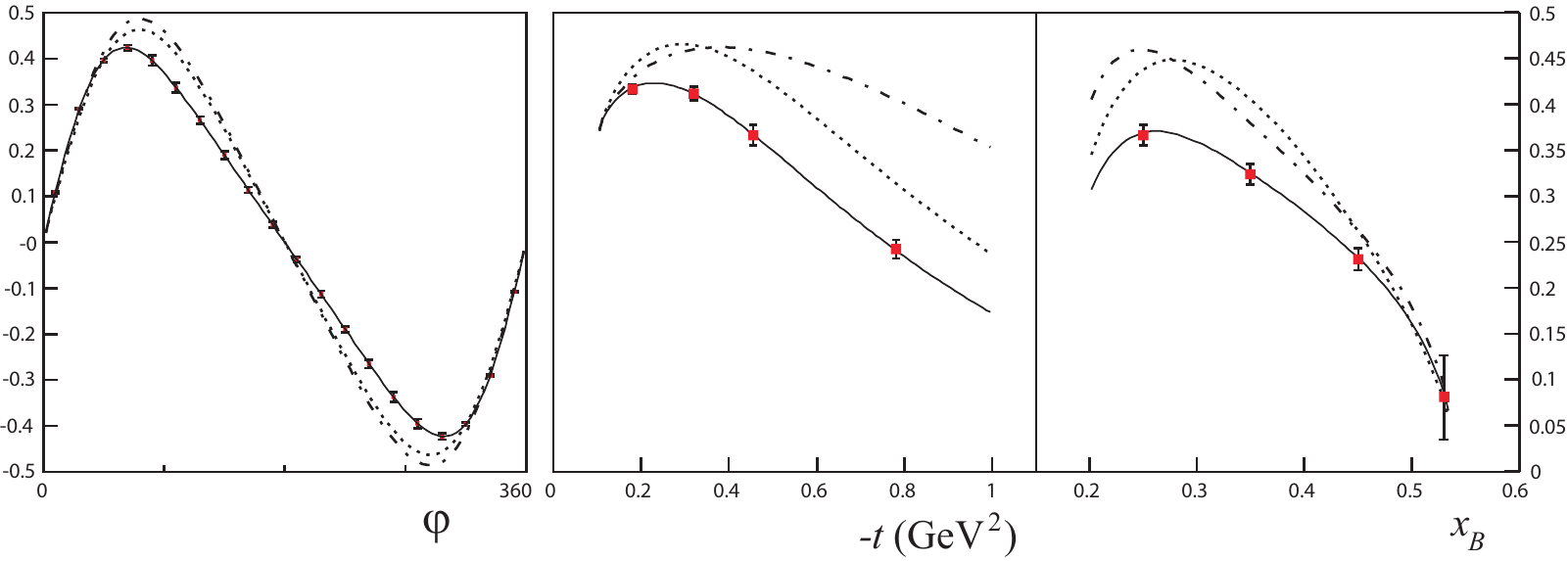}
\caption{The beam spin asymmetry showing the DVCS-BH interference for 11 GeV beam 
energy~\cite{e12-06-119}. Left panel: $x=0.2$, 
$Q^2=3.3$GeV$^2$, $-t=0.45$GeV$^2$. Middle and right panels:  $\phi=90^{\circ}$, 
other parameters same as in left panel. Many other bins will be measured simultaneously. 
The curves represent various parameterizations within the VGG model~\cite{vgg}. 
Projected uncertainties are statistical.}
\label{fig:dvcs_alu_12gev}
\end{figure}

\subsection{Differential cross section for polarized proton target}
The structure of the differential cross section for polarized beam and polarized target 
contains the polarized beam term 
of the previous section and an additional term related to the target polarization~\cite{belitsky02,diehl05}:

\begin{equation}
\sigma_{\vec{e}\vec{p}\rightarrow e\gamma p} = \sigma_{\vec{e}p\rightarrow e\gamma p} +
T[P_\ell\Delta\sigma_{BH} + e_\ell \Delta\tilde\sigma_{INT} + P_\ell e_\ell\Delta\sigma_{INT} + 
\Delta\tilde\sigma_{VCS} + P_\ell\Delta\sigma_{VCS}]
\end{equation}
where the target polarization $T$ can be longitudinal or transverse. If only unpolarized electrons are available, 
the combination 
$-\Delta\tilde\sigma_{INT} + \Delta\tilde\sigma_{VCS}$ can be measured from the differences in the 
target polarizations. If unpolarized electrons and unpolarized positrons are available the combination 
$T\Delta\tilde\sigma_{INT} + \sigma_{INT}$ can be determined at fixed target polarization. 
With both polarized electron and 
polarized positron beams, the combination $T\Delta\tilde\sigma_{INT} + TP_\ell\Delta\sigma_{INT} 
+ P_\ell\tilde\sigma_{INT} + \sigma_{INT}$ can be measured at fixed target polarization. Availability 
of both polarized electron and polarized positron beams thus allows the separation of all contributing terms. 
\begin{figure}[htb]
  \includegraphics[height=.5\textheight]{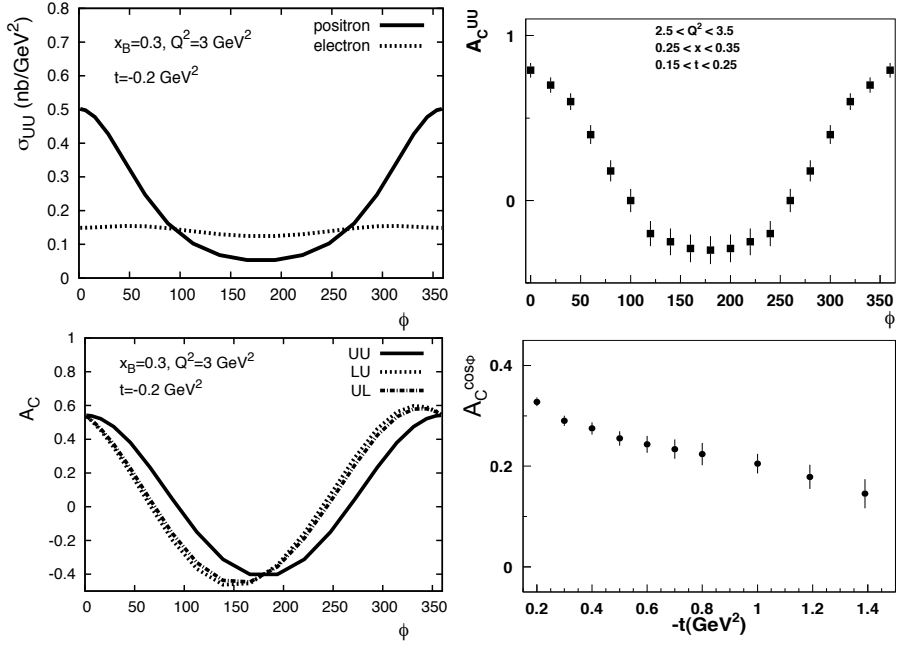}
\caption{Cross sections and charge asymmetries. Top-left: Azimuthal dependence of the unpolarized 
cross section for positron and electron beam at 11 GeV beam 
energy~\cite{guzey}. Bottom-left: Charge asymmetries for unpolarized lepton beams (UU), for leptons 
at fixed polarization (LU), and for protons at fixed polarization (UL). Top-right: Projected statistical 
accuracy for the charge asymmetry of unpolarized lepton beams. 
 Bottom-right: $\cos{\phi}$ moment of the unpolarized charge asymmetry vs momentum transfer $t$ to the 
 proton, for the same $Q^2$ and $x$ bins as the graph at the top.}
\label{fig:cross_section}
\end{figure}
If only polarized electron beams are available a Rosenbluth separation with different beam energies can
separate the term $\Delta\tilde\sigma_{INT}$ from $\Delta\tilde\sigma_{VCS}$, again in a more limited kinematical
range. 
However, the term $\Delta\sigma_{INT}$ can only be determined using the combination of polarized electron 
and polarized positron beams.  
\subsection{Estimates of charge asymmetries for different lepton charges}
For quantitative estimates of the charge differences in the cross sections we use the acceptance and 
luminosity achievable with {\tt\sl CLAS12} as basis for measuring the process 
$ep \rightarrow e\gamma p$ at different beam and target conditions. A 10 cm long liquid hydrogen is
assumed with an electron current of 40nA, corresponding to an operating luminosity of 
$10^{35}$cm$^{-2}$sec$^{-1}$.   For the positron beam a 5 times lower beam current of 8nA is 
assumed. In either case 1000 hours of beam time is used for the rate projections. 
For quantitative estimates of the cross sections the dual model~\cite{guzey} is 
used. It incorporates parameterizations of the GPDs 
$H$ and $E$.  As shown in Fig.~\ref{fig:cross_section}, effects coming from the charge 
asymmetry can be large. In case of unpolarized beam and unpolarized target the cross section
for electron scattering has only a small dependence on azimuthal angle $\phi$, while the corresponding
positron cross section has a large $\phi$ modulation.  The difference is directly related to the 
term $\sigma_{INT}$ in (1). 
\section{Conclusions} 
Availability of a 11 GeV positron beam at the JLab upgrade can significantly enhance the experimental 
DVCS program using {\tt\sl CLAS12} detector in Hall B~\cite{burkert2008}. It allows access to the azimuthally
even BH-DVCS interference terms that are directly related to the real part of the scattering amplitude. 
Moreover, by avoiding use of the Rosenbluth separation technique, the leading contributions to the 
cross sections may be separated in the full kinematical range available at the JLab 12 GeV upgrade.  Even 
at modest positron beam currents of 8nA good statistical accuracy can be achieved for charge 
differences and charge asymmetries. For efficient use of polarized targets higher beam currents of up to
40nA are needed to compensate for the dilution factor of $\sim 0.18$ inherent in the use of currently  
available polarized proton targets based on ammonia as target material, and to allow for a more complete 
DVCS and GPD program at 12 GeV.  

In this talk I have focussed on experiments with large acceptance detectors, which may be the only option given 
the low current expected for positron beams of sufficient good quality. Positron currents in excess of  
$1 \mu A$ are likely going to be required to make such a program attractive for an experimental program
with high resolution magnetic spectrometers.           

\begin{theacknowledgments}
I would like to thank Vadim Guzey for providing me with the predictions of the minimal dual model, and 
Harut Avakian for the experimental projections.  I also like to thank Markus Diehl for providing me with 
slides of his talk on the same subject in Genoa, February 2009, which I used in preparing this talk. 
The Jefferson Science Associates, LLT, operates Jefferson Lab under contract DE-AC05-060R23177. 
\end{theacknowledgments}


\begin{thebibliography}{99}
\bibitem{mdiehl} M. Diehl, Phys. Rept. 388, 41-277, 2003, hep-ph/0307382.
\bibitem{radyush} A.V. Belitsky and A.V. Radyushkin,  Phys. Rept. 418, 1-387, 2005.  
\bibitem{burkardt2003}M. Burkardt, Int. J. Mod. Phys. A18, 173-208, 2003. 
\bibitem{ji2003} X. Ji, Phys. Rev. Lett.91, 062001, 2003. 
\bibitem{step01} S. Stepanyan et al. (CLAS),  Phys. Rev. Lett. 87, 182002, 2001,  arXiv:hep-ex/0107043
\bibitem{girod08} F.X. Girod et al. (CLAS), Phys. Rev. Lett. 100, 162002, 2008, arXiv:0711.4805 [hep-ex]
\bibitem{munoz06} C. Munoz Camacho et al. (Hall A), Phys. Rev. Lett. 97, 262002, 2006,  arXiv:nucl-ex/0607029 
\bibitem{gavalian09} G. Gavalian et al. (CLAS), arXiv:0812.2950 [hep-ex]
\bibitem{chen06} S. Chen et al. (CLAS),  Phys. Rev. Lett. 97, 072002, 2006,  arXiv:hep-ex/0605012
\bibitem{e12-06-119} JLab experiment E12-06-119, F. Sabatie et al. (spokespersons). 
\bibitem{vgg} M. Vanderhaeghen, P. Guichon, M. Guidal,  Phys. Rev. D60, 094017, 1999.
\bibitem{belitsky02} A. Belitsky, D. Mueller, A. Kirchner, hep-ph/0112108, Nucl. Phys. B629:323-392, 2002. 
\bibitem{diehl05} M. Diehl, S. Sapeta, Eur.Phys.J.C41:515-533,2005, hep-ph/0503023 
\bibitem{guzey} V. Guzey and T. Teckentrup, Phys. Rev. D79, 017501 (2009) (see also: http://www.jlab.org/~vguzey/)
\bibitem{burkert2008} V. D. Burkert,  arXiv:0810.4718 [hep-ph].
\end{thebibliography}
\end{document}